\documentclass{aa}
\usepackage{graphics}
\usepackage{latexsym}

\begin{document}
\title{
The cataclysmic variable CW 1045+525:\\ a secondary-dominated dwarf nova?
}
\author{
C. Tappert\inst{1, 2} \and J.R. Thorstensen\inst{3} \and W. H. Fenton\inst{3}
\and N. Bennert\inst{4}
\and L. Schmidtobreick\inst{5, 6} \and A. Bianchini\inst{2}
}
\institute{
Departamento de F\'{\i}sica y Matem\'aticas, Grupo de Astronom\'{\i}a,
Universidad de Concepci\'on, Casilla 160-C, Concepci\'on, Chile 
\and
Dipartimento di Astronomia, Universit\`a di Padova, Vicolo dell'Osservatorio 2,
I-35122 Padova, Italy
\and
Department of Physics and Astronomy, Dartmouth College, Hanover, New Hampshire,
03755 USA
\and
Astronomisches Institut, Ruhr-Universit\"at, D-44780 Bochum, Germany
\and
European Southern Observatory, Casilla 19001, Santiago 19, Chile
\and
Osservatorio Astron\'omico di Padova, Vicolo dell'Osservatorio 5, I-35122 
Padova, Italy
}
\offprints{C. Tappert, \email{claus@gemini.cfm.udec.cl}}
\date{Received / Accepted}
\titlerunning{
The cataclysmic variable CW 1045+525
}
\authorrunning{Tappert et al.}

\abstract{
We present spectroscopic and photometric observations of the cataclysmic
variable CW 1045+525. Both the optical spectrum and the photometric lightcurve
show a strong contribution of a K5V -- M0V secondary. We derive an orbital
period $P_{\rm orb}$ = 0.271278(1) d by measuring the radial velocities of
the absorption lines of the secondary. The period and spectral type of the
secondary suggest a distance of 350 -- 700 pc. There is evidence for additional
sources of line- and continuum emission, but no direct evidence of an accretion
disc. We discuss several scenarios for the nature of CW 1045+525 on the basis 
of our results, finding a dwarf nova classification the most probable, although
not completely satisfying, explanation for the observed characteristics.  
}

\maketitle

\section{Introduction}

Cataclysmic variables (CVs) are close binaries consisting of a white dwarf 
primary and usually a late-type main-sequence secondary. The latter fills its 
Roche lobe and thus enables mass transfer into the gravitational regime of the
primary. Due to the conservation of angular momentum, this transfer -- in the
absence of strong magnetic fields -- takes place via an accretion disc.
For a comprehensive overview on CVs see Warner (\cite{warn95}).

The little studied object CW 1045+525 was discovered as part of the Case 
Low-Dispersion Northern Sky Survey. It showed strong Balmer emission lines 
superposed on a late-type spectrum, but with a very strong ultraviolet 
continuum. Pesch \& Sanduleak (\cite{pescsand87}) drew attention to the object,
noting that the spectrum was quite unusual. Wagner et al.\ (\cite{wagn+88}) and
Szkody \& Howell (\cite{szkohowe92}) both published spectra which show the 
broad Balmer and HeI emission characteristic of cataclysmic binaries, together 
with very strong K-star features.  The object is listed as `UMa5' in the Downes
\& Shara (\cite{downshar93}) atlas. Henden \& Honeycutt (\cite{hendhone95}) 
published a photometric sequence for the field. Although the system is 
generally classified as a `dwarf nova', we were unable to find any indication 
that outbursts have been observed.

We here present photometric time series and time-resolved spectroscopy of
CW 1045+525. Preliminary evaluations of part of this data have been presented
by Bennert et al.\ (\cite{benn+99b}) and Tappert et al.\ (\cite{tapp+00}).

\section{Observations and reduction}

\begin{table*}
\caption[]{Overview on the observations. $n_{\rm data}$ gives the number of
data points, $t_{\rm exp}$ the individual exposure time, $\Delta t$ the time
range covered by the observations, $\Delta \lambda$ the FWHM resolution, and
$\Delta V$ ($\Delta B$) the magnitude range of the object during the night
in $V$ ($B$). For the 2.4m Hiltner observations, $\Delta \lambda$ was measured
by fits to the night sky line, the corresponding data for the 1.82m Ekar refer
to fits to the calibration spectra. The magnitudes for the Hoher List and the
Ekar data have been derived by differential photometry and comparison to
secondary photometric standards, and should be accurate within the cited
errors.}
\label{obs_tab}
\begin{tabular}{l c r r l c l c}
\hline
date & telescope & $n_{\rm data}$ & $t_{\rm exp}$ [s] & $\Delta t$ [h] 
& filter / range [{\AA}] & $\Delta \lambda$ [{\AA}] & $\Delta V$ / $\Delta B$
[mag] \\
\hline
1998-03-12 & 1m Hoher List & 92 & 90 & 2.78 & V & -- 
& 15.399(35) -- 15.191(53) \\
1999-03-10 & 1m Hoher List & 150 & 120 & 6.74 & V & -- 
& 15.439(21) -- 15.165(20) \\
1999-03-11 & 1m Hoher List & 191 & 120 & 8.28 & V & -- 
& 15.408(23) -- 15.109(17) \\
1999-03-12 & 1m Hoher List & 18 & 120 & 1.39 & V & -- 
& 15.327(29) -- 15.108(73) \\
1999-03-13 & 1m Hoher List & 58 & 120 & 6.72 & V & -- 
& 15.62(14) -- 15.02(23) \\
1999-03-14 & 1m Hoher List & 163 & 120 & 8.47 & V & -- 
& 15.358(22) -- 15.094(20) \\
2000-01-07 & 2.4m Hiltner & 6 & 360 & 2.61 & 4210 -- 7560 & 3.6 \\
2000-01-08 & 2.4m Hiltner & 5 & 360 & 7.08 & 4210 -- 7560 & 3.6 \\
2000-01-09 & 2.4m Hiltner & 3 & 360 & 5.86 & 4210 -- 7560 & 3.6 \\
2000-01-10 & 2.4m Hiltner & 7 & 360 & 0.65 & 4210 -- 7560 & 3.6 \\
2000-03-02 & 1.82m Ekar & 2 & 60 & 0.26 & B & -- & 16.005(26) -- 15.997(26) \\
2000-03-02 & 1.82m Ekar & 17 & 60 & 2.50 & V & -- & 15.197(19) -- 15.083(19) \\
2000-03-02 & 1.82m Ekar & 2 & 60 & 0.24 & R & -- & \\
2000-03-02 & 1.82m Ekar & 1 & 300 & -- & 4444 -- 6677 & 9.2 & \\
2000-03-03 & 1.82m Ekar & 8 & 60 & 5.50 & B & -- & 15.955(26) -- 15.685(26) \\
2000-03-03 & 1.82m Ekar & 9 & 30 & 5.50 & V & -- & 15.163(19) -- 14.945(20) \\
2000-03-03 & 1.82m Ekar & 9 & 30 & 5.47 & R & -- & \\
2000-03-03 & 1.82m Ekar & 18 & 600 & 6.39 & 6328 -- 8430 & 8.4 & \\
2000-04-07 & 1.82m Ekar & 7 & 20 & 6.79 & V & -- & 15.156(28) -- 15.008(26) \\
2000-04-07 & 1.82m Ekar & 16 & 300 & 7.21 & 4460 -- 6660 & 9.2 & \\
2000-04-07 & 1.82m Ekar & 13 & 300 & 7.17 & 6340 -- 8442 & 8.4 & \\
2001-03-26 & 2.4m Hiltner & 1 & 480 & -- & 4210 -- 7560 & 3.8 \\
2001-03-27 & 2.4m Hiltner & 1 & 480 & -- & 4210 -- 7560 & 3.8 \\
2001-05-12 & 2.4m Hiltner & 2 & 480 & 1.64 & 4210 -- 7560 & 5.9 \\ 
2001-05-14 & 2.4m Hiltner & 1 & 480 & -- & 4210 -- 7560 & 3.7 \\
2001-05-17 & 2.4m Hiltner & 1 & 480 & -- & 4210 -- 7560 & 3.7 \\
\hline 
\end{tabular}
\end{table*}

The 1998 and 1999 data were obtained during the Astro\-no\-misches 
Beobachtungspraktikum (e.g., Schmidtobreick et al.\ \cite{schm+00}) at the 
1m Cassegrain reflector of the Hoher List Observatory\footnote{The Hoher List 
Observatory is operated by the University of Bonn}, Germany. The telescope was 
equipped with a Ford Loral FA2048 CCD and a Johnson V filter. The CCD is read 
out through two different channels, thus minimizing read-out time, but making a
separate bias correction for both halfs of the CCD necessary. Apart from that,
the reduction was done in the standard way, using IRAF\footnote{IRAF is 
distributed by the National Optical Astronomy Observatories} tasks to apply 
overscan and domeflats. 

The 2000 January observations were taken by JRT using the 2.4-m Hiltner 
telescope at MDM Observatory, Kitt Peak, Arizona, USA, with the modular 
spectrograph, and the same equipment and protocols were used by WHF for
the 2001 March and 2001 May observations. For the MDM observations 
we observed comparison lamps frequently,
and checks of the $\lambda 5577$ night-sky line show that our wavelength scale 
is typically stable to $\sim$5 km s$^{-1}$.  When the weather appeared 
photometric we observed flux standards in twilight. Even so, our absolute 
fluxes are not expected to be accurate to much better than 30 per cent, because
of occasional clouds and variable losses at the spectrograph slits ($1''$ at 
the 2.4 m). Furthermore, for unknown reasons the modular spectrograph produces 
wavelike distortions in the continua; these appear to average out in sums of 
many exposures.

The MDM data were reduced to counts vs. wavelength at the observatory, using 
standard IRAF procedures. Radial velocities of emission lines were measured 
using convolution algorithms described by Schneider \& Young 
(\cite{schnyoun80}) and Shafter (\cite{shaf83a}). For absorption velocities,
we used the cross-correlation algorithms of Tonry \& Davis (\cite{tonrdavi79}),
as implemented in the {\it rvsao} package (Kurtz \& Mink \cite{kurtmink98}).
To search for periods we used the `residual-gram' method described by
Thorstensen et al.\ (\cite{thor+96}), and to test alias choices in doubtful
cases we used the Monte Carlo method explained by Thorstensen \& Freed
(\cite{thorfree85}).

In March and April 2000 the system was observed with the 1.82 m telescope at
Mt. Ekar, Asiago Observatory, Italy. The AFOSC\footnote{Asiago Faint Object
Spectrograph and Camera} system was used, thus enabling both spectroscopic and
photometric measurements without changing the instrumental configuration. In
March we used a red grism (\#8) for time-resolved spectroscopy and B, V, R 
filters for accompanying time-series photometry. The sequence of the 
measurements was chosen to be spectrum -- B -- V -- R, in an attempt to 
compare 
radial velocities with the lightcurve, and to gather information on the 
different components which contribute to the latter. However, poor weather 
conditions causing long integration times and frequent gaps spoiled that 
effort. For the April observations we therefore chose to concentrate on the 
spectroscopy and opted for a sequence of four spectroscopic measurements 
followed by one image in the V passband. The spectroscopy was done by 
alternating between a blue (\#7) and a red grism (\#8). Every measurement was 
followed by corresponding wavelength calibration exposures (a He lamp for grism
\#7, Ne for grism \#8). Like in March, the weather was not photometric, so that
we refrained from observing flux standards.

The radial velocities of the March and April data were measured with respect
to an average of several night-sky lines, by fitting single Gauss functions
to the lines both from the object and from the night sky. The period search 
within the individual data sets were done by using the Scargle (\cite{scar82})
and AOV (Schwarzenberg-Czerny \cite{schw89}) algorithms implemented in
ESO-MIDAS. A method similar to that of Thorstensen \& Freed (\cite{thorfree85})
was used to check for alias periods (Mennickent \& Tappert \cite{menntapp01}).

The instrumental magnitudes of the photometric data were extracted with respect
to an average lightcurve composed out of several comparison stars. Three of 
these stars were included in the atlas of secondary photometric standards by 
Henden \& Honeycutt (\cite{hendhone95}), enabling the computation of calibrated
magnitudes for CW 1045+525 within an error of $\sim$0.02 mag.

Table \ref{obs_tab} summarizes the data on the observations.

\section{Results}

\subsection{Spectrum and orbital period}

\subsubsection{Properties of the average spectrum}

\begin{figure}
\resizebox{8.5cm}{!}{\includegraphics{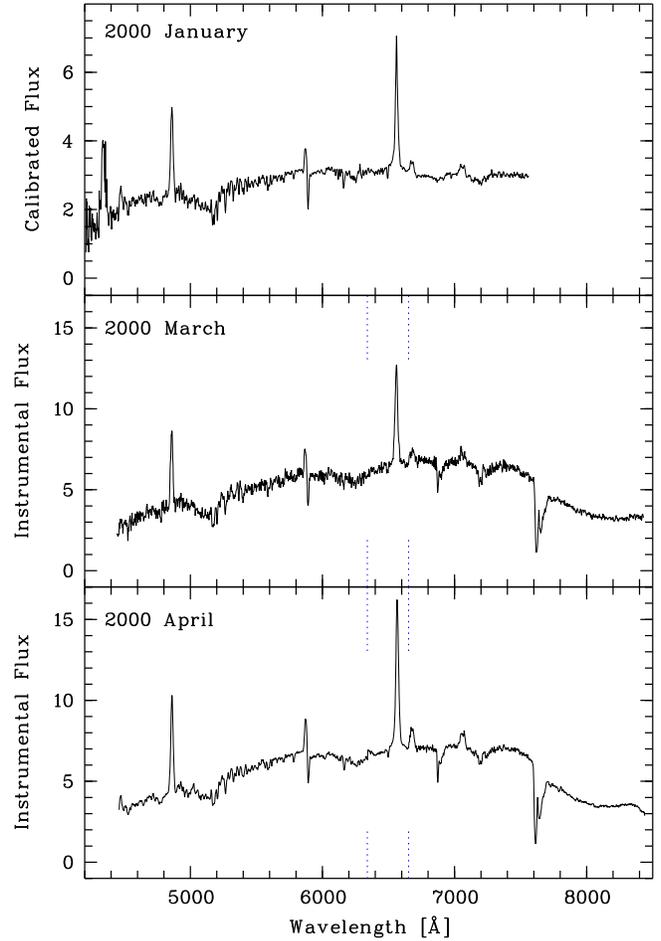}}
\caption[]{Average spectra of CW1045+525 from January (top), March (middle),
and April (bottom), 2000. The dotted lines in the lower plots mark the 
overlapping region of grism \#7 and grism \#8. The flux of the upper plot is
in $10^{-15}$ erg cm$^{-2}$ s$^{-1}$ {\AA}$^{-1}$, the instrumental flux of the
two lower ones has been divided by 100 for presentation purposes. Note that 
those spectra have not been corrected for the instrumental response curve.}
\label{avsp_fig}
\end{figure}

\begin{table}
\caption[]{Properties of the principle spectral features of the average 
spectra. Negative equivalent widths $W_{\lambda}$ denote emission lines. Colons
mark uncertain data due to noise or blends.}
\label{avsp_tab}
\begin{tabular}{l l l r r}
\hline
$\lambda$ [{\AA}] & ID & $\lambda_{\rm rest}$ & $W_{\lambda}$ [{\AA}] 
& FWHM [{\AA}] \\
\hline
\multicolumn{5}{l}{\sl 2000 January:} \\
4340 & H$\gamma$ & 4341 & $-$43 & 25: \\
4473 & \ion{He}{I} & 4471 & $-$7 & 19: \\
4860 & H$\beta$ & 4861 & $-$25 & 21 \\
5871 & \ion{He}{I} & 5876 & $-$5: & 17: \\
5892 & \ion{Na}{I} D & 5890/5896 & 4: & 11: \\
6561 & H$\alpha$ & 6563 & $-$26 & 22 \\
6678 & \ion{He}{I} & 6678 & $-$4 & 30 \\
7058 & \ion{He}{I} & 7065 & $-$4 & 45: \\
\multicolumn{5}{l}{\sl 2000 March:} \\
4859 & H$\beta$ & 4861 & $-$22 & 20 \\
5869 & \ion{He}{I} & 5876 & $-$5: & 15: \\
5892 & \ion{Na}{I} D & 5890/5896 & 4: & 12: \\
6559 & H$\alpha$ & 6563 & $-$24 & 25 \\
6682 & \ion{He}{I} & 6678 & $-$5 & 40 \\
7056 & \ion{He}{I} & 7065 & $-$4 & 44 \\
\multicolumn{5}{l}{\sl 2000 April:} \\
4473 & \ion{He}{I} & 4471 & $-$5: & 15: \\
4861 & H$\beta$ & 4861 & $-$33 & 22 \\
4928 & \ion{He}{I} & 4922 & $-$6: & 11: \\
5022 & \ion{He}{I} & 5016 & $-$3: & 27: \\
5872 & \ion{He}{I} & 5876 & $-$6: & 17: \\
5894 & \ion{Na}{I} D & 5890/5896 & 3: & 11: \\
6566 & H$\alpha$ & 6563 & $-$34 & 24 \\
6678 & \ion{He}{I} & 6678 & $-$6 & 35 \\
7059 & \ion{He}{I} & 7065 & $-$7 & 43 \\
\hline 
\end{tabular}
\end{table}

\begin{figure}
\resizebox{8cm}{!}{\includegraphics{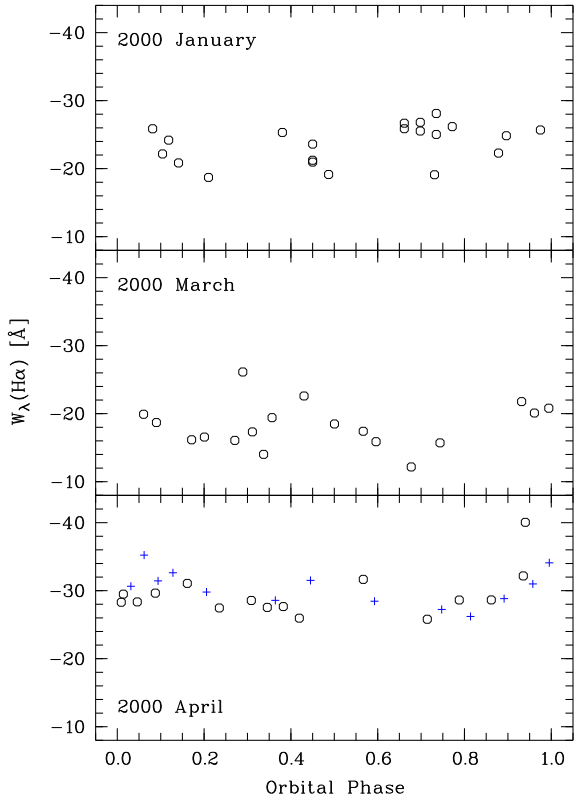}}
\caption[]{Equivalent widths of the H$\alpha$ emission line from January
(top), March (middle), and April (bottom), data. In the latter plot, the
different symbols denote measurements with grism \#7 ($+$) and grism \#8
({\Large $\circ$}).}
\label{eqws_fig}
\end{figure}

The averaged spectra for January, March, and April 2000, are shown in 
Fig.\ 
\ref{avsp_fig}. In the case of the latter two, first the individual spectra of 
grism \#7 and of grism \#8 were averaged. The resulting spectra were rebinned
to the dispersion of grism \#8, and, after ensuring that the overlapping 
regions would match, were then combined to yield a single average spectrum over
the whole spectral range. Although these spectra are not flux-calibrated, the 
comparison with the Kitt Peak data shows that the instrumental response curve
affects the continuum shape only weakly.

Table \ref{avsp_tab} shows the properties of the principle spectral features
of the average spectra, i.e.\ of the Balmer and \ion{He}{I} emission, and of 
the \ion{Na}{I} absorption. Noteworthy is the total absence of high ionized 
emission lines like \ion{He}{II} $\lambda$4686.

Comparing the average spectra for each month shows that while the March and 
January data have very similar values, the equivalent widths $W_{\lambda}$ of 
all emission lines in the April data are significantly higher. However, the
calibrated magnitudes in Table \ref{obs_tab} show no corresponding behaviour
for the continuum for the March and April data (the uncertainties of
spectrophotometrically derived magnitudes of the January data impede a 
conclusive comparison). It appears that the difference is due to an
enhancement of the line-emission region only.

In order to check if this phenomenon corresponds to real long-term behaviour,
or if it is due to orbital changes and average spectra composed from data not
uniformly distributed in phase, we measured $W_{\lambda}$ for the H$\alpha$
emission line of the individual spectra and folded it according to the 
ephemeris derived below. The result in Fig.\ \ref{eqws_fig} clearly shows a) 
that there is no convincing phase-dependent variation of $W_{\lambda}$, and b) 
that on average there are significant differences for the individual data sets,
probably reflecting changes in the mass-transfer rate. This behaviour is also
evident in the traced spectrum of the H$\alpha$ line (Fig.\ \ref{trcsp_fig}).

The spectrum of CW 1045+525 (Fig.\ \ref{avsp_fig}) shows a strong late-type 
contribution, similar to previously-published spectra. We estimated the 
secondary's contribution and spectral type using K dwarf spectra from Pickles'
(\cite{pick98}) library.  We scaled these with a range of factors, and 
subtracted them from a version of our mean spectrum which had been smoothed
to match the somewhat lower resolution of the library spectra. Acceptable
cancellation of the late-type features (especially the broad absorption feature
centered near $\lambda 5180$) could be obtained for stars of type K5 through 
M0, but only if the subtracted signal constituted most of the light in the 
observed range. If we assume the same brightness level of the system during all
spectroscopic observations, we obtain a mean magnitude for the average Januray
spectrum of $V \sim 15.1$; in the best deconvolution, about 70 per cent of the 
flux  is from the K-star, yielding an average $V \sim 15.5$ for the K star 
alone.

\subsubsection{Radial velocities and the orbital period\label{radvecsec}}

\begin{figure}
\rotatebox{270}{\resizebox{5.5cm}{!}{\includegraphics{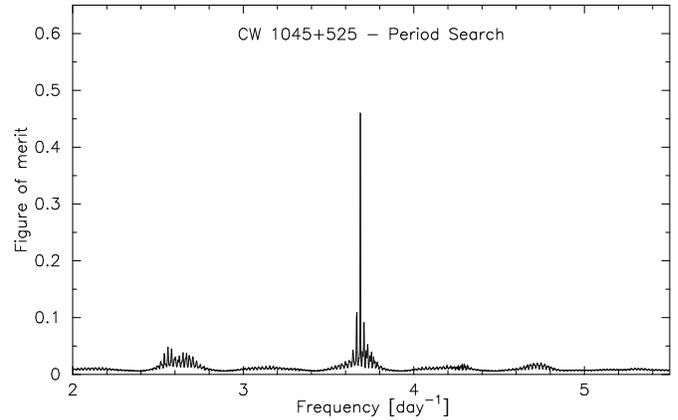}}}
\caption[]{Period search of the absorption-line radial velocities in the
vicinity of the best fit.
The figure of merit is $1./[\sum_i((o_i - c_i)/\sigma_i)^2]$, where $o_i$ are the 
observed velocities, $c_i$ the velocities computed from the best-fitting
sinusoid at each trial frequency, and $\sigma_i$ are the estimated 
velocity uncertainties.  
}
\label{psrch_fig}
\end{figure}

\begin{figure}
\rotatebox{270}{\resizebox{5.5cm}{!}{\includegraphics{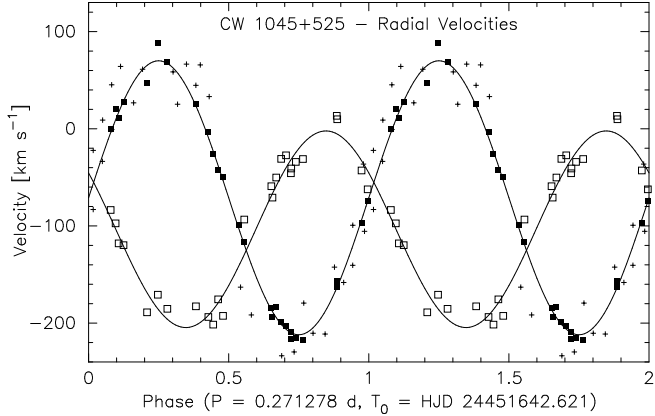}}}
\caption[]{Radial velocity data folded with 
$P$ = 0.271278 d with respect to the zero point (inferior conjunction) of the 
MDM absorption spectra (filled squares). The small crosses show the 
2000 April Ekar absorption velocities, offset by $-63$ km s$^{-1}$ to
match the MDM zero point.  H$\alpha$ emission velocities
are shown by 
open squares. Also shown are the best sinuoids for both components. Note the
evident phase shift $\Delta \varphi$ = 0.595(10) cycles of the sinusoids.}
\label{rv_fig}
\end{figure}

\begin{table*}
\caption[]{Parameters of the sine fits to the MDM radial velocities.
Fits are of the form $v(t) = \gamma + K \sin [2 \pi (t - T_0) / P].$
For the fit to the H$\alpha$ velocities, $P$ was held fixed.}
\begin{tabular}{l l l l l l}
\hline
data set & $T_0$ [HJD] & $P$ [d] & $K$ [km s$^{-1}$] & 
$\gamma$ [km s$^{-1}$] & $\sigma$ [km s$^{-1}$] \\
\hline
Absn. & 2451553.913(1) & 0.271278(1) & 141(3) & $-71$(2) & 7 \\
H$\alpha$ & 2451553.804(2) & [0.271278] & 101(5) & $-103$(3) & 14 \\
\hline
\end{tabular}
\label{mdmvfit_tab}
\end{table*}

For the MDM data, we measured radial velocities of the late-type spectrum 
using cross-correlation techniques. For the template spectrum we used a 
velocity-compensated sum of spectra of G- and K-type IAU velocity standards. To
avoid emission lines only the region $5020 < \lambda <  5800$ {\AA} was used. 
The correlation peaks were very strong and the estimated statistical 
uncertainties were generally $< 10$ km s$^{-1}$. The H$\alpha$ emission line 
was measured using a correlation function consisting of positive and negative 
Gaussians, each with FWHM 8 {\AA} and separated by 30 {\AA}. 
A period near 0.27 d was plainly apparent in both the absorption and
emission-line velocities.  The Ekar absorption
velocities (2000 April) are less precise than those from MDM but
they do show the modulation, and help constrain the ephemeris.  
They were included in the period search analysis, with a $-63$ km s$^{-1}$
zero point offset to bring them in line with the MDM velocities.  

Fig. \ref{psrch_fig} shows a period search of the combined absorption
velocity data, which span 495 d of elapsed time.  Because the uncertainties
in the velocities are much smaller than their modulation, a single
frequency corresponding to $P_{\rm orb}$ = 0.271278 d is selected 
without any cycle-count ambiguity.  Fig. \ref{rv_fig} shows the
emission and absorption velocities folded on this period, with the
best fitting sinusoids superposed. The blue-to-red
crossing of the absorption velocities follow the ephemeris 
\begin{equation}
T_0 = {\rm HJD}~2451642.621(1) + 0.271278(1) E,
\label{ephem_eq}
\end{equation}
where $E$ is an integer.  The parameters of the best-fit sinusoids are
given in Table \ref{mdmvfit_tab}.  

The absorption-line velocities should rather faithfully reflect the {\it phase} 
of the secondary star's motion (though the {\it amplitude} may be affected by 
illuminated-atmosphere and other effects; Wade \& Horne \cite{wadehorn88}). 
If the emission-line velocities were to follow the motion of the white dwarf,
the phase of their motion should be 1/2 cycle from the absorption line phase.
It is therefore sobering that the emission and absorption line phases differ by
significantly more, specifically $0.595  \pm 0.010$ cycles. Furthermore, the
mean velocities $\gamma$ differ significantly. These differences are clear in 
Fig.\ \ref{rv_fig}. This underscores the dangers of using emission lines
for dynamical purposes -- while they reliably indicate the period, they may not
reflect the center-of-mass motion of either star (e.g., Tappert \cite{tapp99}).

\subsubsection{Line profile analysis}

\begin{figure}
\resizebox{9cm}{!}{\includegraphics{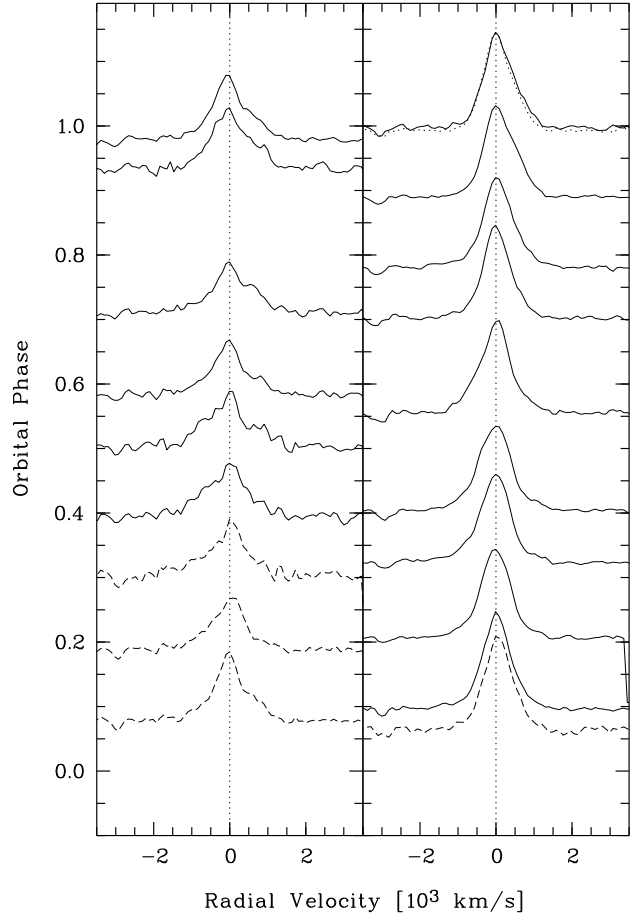}}
\caption[]{Traced spectrum of the H$\alpha$ emission line in the Ekar spectra 
(right: 2000-03-03, left: 2000-04-07; grism \#8 only), subsequently binned into 
10\% phase bins. The time sequence is symbolized by the line styles dotted -- 
solid -- dashed. Zero phase refers to the ephemeris (\ref{ephem_eq}). Note the
asymmetric line profile especially evident in phases 0.6 and 1.0.}
\label{trcsp_fig}
\end{figure}

\begin{figure}
\resizebox{8cm}{!}{\includegraphics{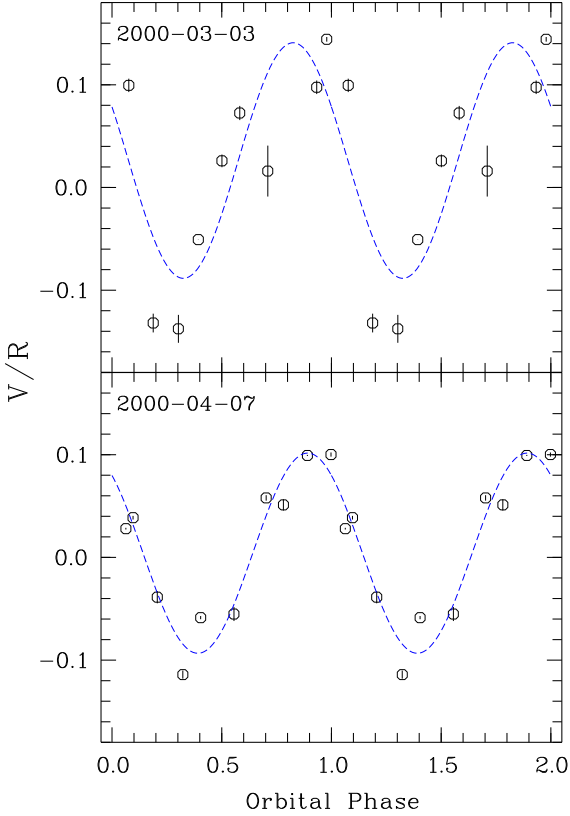}}
\caption[]{$V/R$ plot of the H$\alpha$ line from the March (top) and the
April (bottom) data. The dashed curves give the best sine fits to the 
individual
data sets.}
\label{vr_fig}
\end{figure}

\begin{table}
\caption[]{Parameters of the sine fits to the $V/R$ values of the 
H$\alpha$ 
line. The zero point $\varphi_0$ refers to the negative-positive crossing of
the fit and corresponds to the superior conjunction of the emission source.}
\begin{tabular}{l l l l l}
\hline
data set & $\gamma$ & $K$ & $\varphi_0$ & $\sigma(K)/K$ \\
\hline
03-03 & 0.026(13) & 0.115(21) & 0.575(28) & 0.183 \\
04-07 & 0.004(05) & 0.097(07) & 0.641(14) & 0.067 \\
\hline
\end{tabular}
\label{vrfit_tab}
\end{table}

\begin{figure}
\resizebox{9cm}{!}{\includegraphics{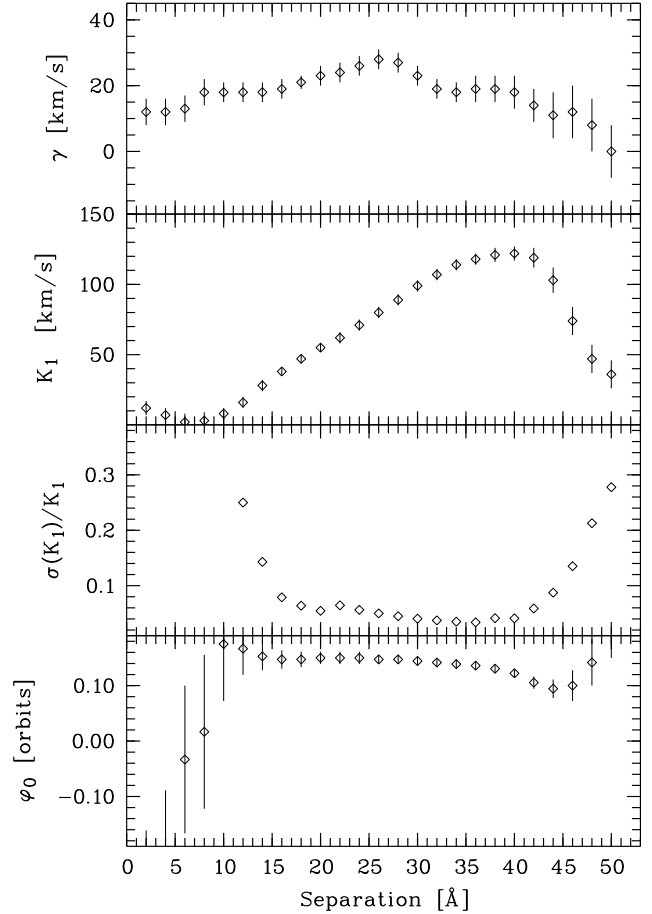}}
\caption[]{Diagnostic diagram of the 2000-04-07 data, showing the parameters
of the radial-velocity fit $v_r (\varphi) = \gamma - K_1 \sin (2 \pi \varphi)$
as functions of the Gaussian separation. The zero
point of the orbital phase, $\varphi_0 = 0.0$, refers to ephemeris 
(\ref{ephem_eq}). Note that this point is not reached for separations which
correspond to data not dominated by noise (separations $d < 44$ {\AA}).}
\label{dd_fig}
\end{figure} 

The differences in phase between the absorption and the emission lines found
in the last section indicate the presence of an isolated emission source, 
i.e.\ an emission region which is not distributed symmetrically around the
white dwarf. A distribution of the latter type, e.g.\ an emission ring within
the accretion disk, would result in a symmetric line profile, with orbital
variations only in position, but not in shape (e.g., Horne \& Marsh 
\cite{hornmars86}, and references therein). In the case of CW 1045+525,
however, the line profile of H$\alpha$ is clearly asymmetric at certain 
phases (Fig.\ \ref{trcsp_fig}).

In order to obtain a quantitative description of this behaviour, one can 
compare the flux in the blue half $F(V)$ of the line profile to the one in the 
red half $F(R)$ by defining (Tappert \cite{tapp99})
\begin{equation}
 V/R = \log \frac{F(V)}{F(R)}
 \label{vrdiff_eq}
\end{equation}

The point which separates both halves is here chosen as the centre of the line 
flanks at three intensity values, specifically at 1/4, 1/5, and 1/6, of the 
maximum intensity. This proved to be sufficiently above the noise-dominated 
extreme line wings, as well as far enough below the peak to consider the main 
part of the line. The actual $V/R$ is then taken as the average of the three 
values, its sigma being indicative of the noise in the line. A more detailed 
description of this method will be given elsewhere (Tappert et al., in 
preparation).

The resulting $V/R~(\varphi)$ were fitted with a sine function
\begin{equation}
V/R~(\varphi) = \gamma - K \sin (2 \pi \varphi). 
\label{sine_eq}
\end{equation}
The  derived parameters are listed in Table \ref{vrfit_tab}, the actual
fits and the data are plotted in Fig.\ \ref{vr_fig}. The relative
amplitude error $\sigma(K)/K$ is a measure for the significance of the
variation. As a comparison, a corresponding examination of a symmetric, 
undisturbed, line profile gave $\sigma(K)/K$ = 0.66 (Tappert et al., in 
preparation). Therefore, CW 1045+525 shows a significant variation in both
data sets. The sigma in March is higher than in April, due to the lower S/N of
the data. Nevertheless, both data sets show basically identical variations, 
with zero points of the negative-to-positive crossing at $\varphi_0 \sim 
0.6$. This point corresponds to the superior conjunction of the additional
emission source (the asymmetric emission changes from the red to the blue 
half).\footnote{Note that the corresponding radial-velocity curve is expected 
to be antiphased to the $V/R$ curve. This is not the case for the 
data presented here, as our radial-velocity measurements refer to the 
wings of the 
line profile.} The obtained location in our data is thus a rather unusual one. 
Emission from the bright spot region or from the secondary, e.g., would have 
values of $\varphi_0 \sim 0.35$ and 0.5, respectively.

The spectrum in Fig.\ \ref{trcsp_fig} shows furthermore, that the line
peak has rather low radial velocity and stays close to the dotted line which
marks zero radial velocity, while the line wings seem to have higher values. 
To investigate this in more detail, we computed a diagnostic diagram
(Shafter \cite{shaf83a}). This method shows the parameters of the 
radial-velocity curve as a function of the separation of the two Gaussians 
used for the velocity determination (see Section \ref{radvecsec}). Figure
\ref{dd_fig} shows the result of using Gaussians with FWHM = 4 {\AA} with the
April data. The fact that the semi-amplitude $K_1$ increases with the 
separation confirms our suspicion based on the visual impression. It appears
thus that the line profile of CW 1045+525 consists of a low-velocity, almost 
stationary, central component and ``high-velocity''\footnote{We here use
quotations marks, as these velocities stay rather within the range of the
usually observed values for CVs, and to distinguish them from the high-velocity
components observed in certain nova-likes (e.g., Taylor et al.\ \cite{tayl+99},
for \object{LS Peg}).} wings. With this finding, the precise phasing of the 
$V/R$ variation loses importance, as it is not mainly caused by the movement of
an additional emission (like e.g.\ in \object{U Gem}; Stover \cite{stov81a}), 
as rather by the movement of the line wings around the (almost) stationary 
component, explaining the phase shift of roughly 0.5 orbits between the $V/R$ 
zero phase and the line wings.

\subsection{Lightcurve}

\begin{figure}
\resizebox{8.1cm}{!}{\includegraphics{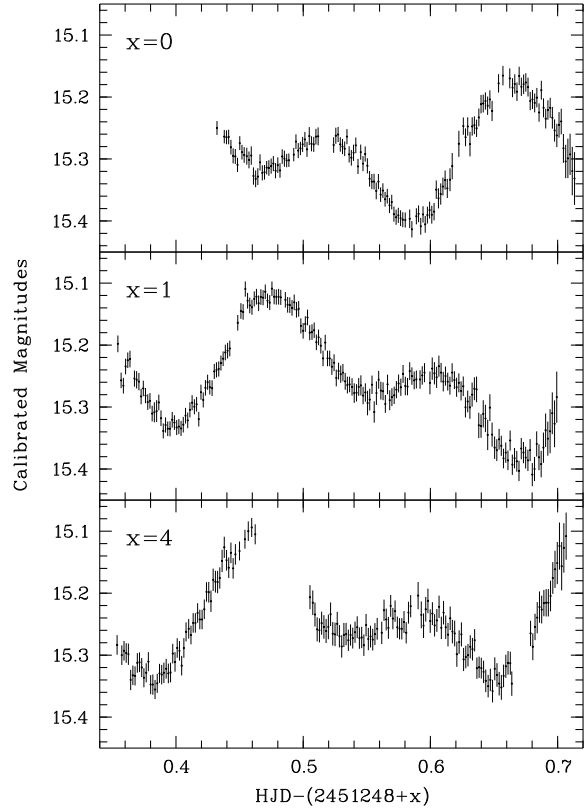}}
\caption[]{Individual lightcurves from 1999-03-10 (top), 11 (middle), and 14
(bottom).}
\label{ind_lcs_fig}
\end{figure}

\begin{figure}
\rotatebox{270}{\resizebox{6.3cm}{!}{\includegraphics{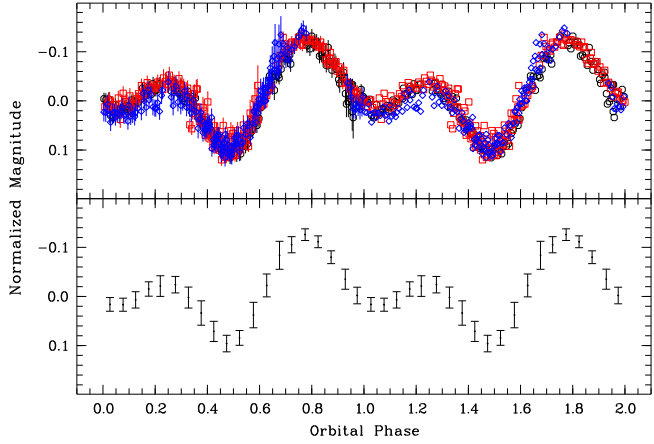}}}
\caption[]{Orbital lightcurve of CW 1045+525. {\bf Top:} Combined normalized
data from 1999-03-10 ({\Large $\circ$}), 11 ($\Box$), and 14 ($\Diamond$).
A linear decline with a slope of 0.31 mag/d has been subtracted from the
1999-03-11 data. For presentation purposes, two orbital phases are shown, the 
second one without error bars. {\bf Bottom:} The data from the upper plot 
averaged into 0.05 phase bins. The error bars indicate the sigma of the 
averaging process.}
\label{avlc_fig}
\end{figure}

\begin{table}
\caption[]{Description of the features in the average lightcurve and their
possible explanation. The phase error was estimated on the basis of several 
measurements with a graphics cursor.}
\begin{tabular}{l l l}
\hline
feature & phase & explanation \\
\hline
shallow minimum & 0.05(1) &
\begin{minipage}[t]{4.05cm}
inferior conjunction of the secondary, minimum surface brightness due to
circular projection, additional source unobscured
\end{minipage} \\
\noalign{\smallskip}
small maximum & 0.24(1) &
\begin{minipage}[t]{4.05cm}
secondary at 90\degr, maximum surface brightness due to maximally projected 
elongation, additional source (partly) obscured
\end{minipage} \\
\noalign{\smallskip}
deep minimum & 0.49(1) &
\begin{minipage}[t]{4.05cm}
superior conjunction of the secondary, minimum surface brightness due to
circular projection, additional source (partly) obscured
\end{minipage} \\
\noalign{\smallskip}
large maximum & 0.77(1) &
\begin{minipage}[t]{4.05cm}
secondary at 270\degr, maxi\-mum surface brightness due to maximally projected
elongation, additional source unobscured
\end{minipage} \\
\noalign{\smallskip}\hline
\end{tabular}
\label{lc_tab}
\end{table}

\begin{figure}
\resizebox{8.5cm}{!}{\includegraphics{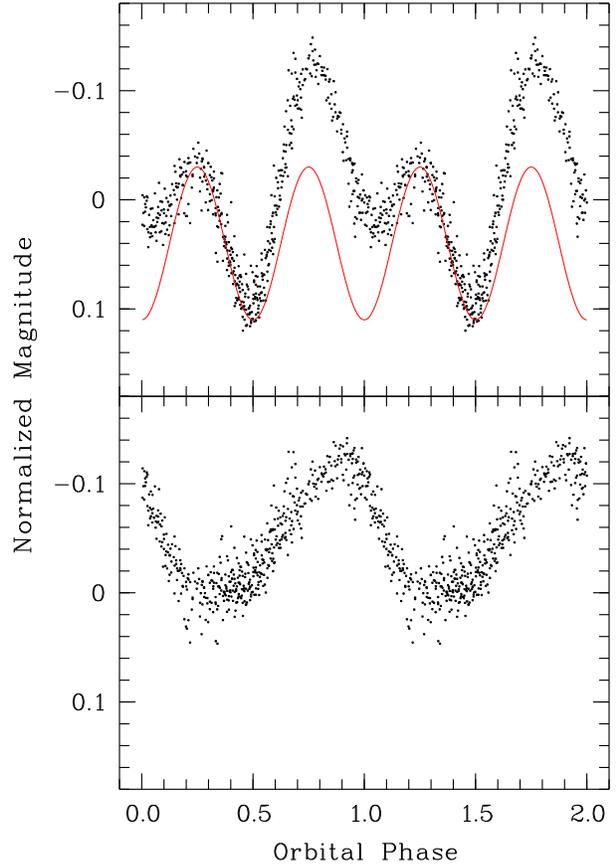}}
\caption[]{The combined lightcurve, together with a sinusoidal function shifted
by 0.25 orbital phases and adjusted in zero point and amplitude to match the
data (top), and the sine-subtracted data (bottom).}
\label{lcsine_fig}
\end{figure}

Of all the photometric data, only the measurements at the Hoher List from 
1999-03-10, 11, and 14, have both sufficient time resolution and coverage to 
define the shape of the lightcurve. 
These individual data sets are presented
in Fig.\ \ref{ind_lcs_fig}. Each set covers a time range larger than the
orbital period (Table \ref{obs_tab}). We note that both in the middle and the 
lower plot the distance between the two deep minima amounts to exactly one 
orbital cycle. In the upper data set, this time range roughly corresponds to
the first, shallow, minimum and the end of the data set. As the individual
lightcurves are furthermore of very similar appearance, it seems safe to
assume that the observed variation with $\Delta V \sim 0.3$ mag is modulated 
with the orbital period.

Apart from this variation, which will be discussed below, the plot sequence 
also indicates some minor variations at non-orbital time-scales, especially 
evident in the data from 1999-03-11 (middle plot in Fig.\ \ref{ind_lcs_fig}) 
which inhabits a linear decline of $\sim 0.3$ mag/d. We furthermore note that
all lightcurves are remarkably smooth and do not show any clear evidence
for flickering, usually the signature of an accretion disc in CVs. This
behaviour, together with the distinctive shape of the lightcurve, is very
reminiscent of a magnetic CV in low state (e.g., Burwitz et al.\ 
\cite{burw+98}). We discuss the pro and cons of this possibility in Section 
\ref{polar_sec}.

The spectroscopically derived ephemeris (\ref{ephem_eq}) allows an
extrapolation to the photometric measurements with a precision of 0.005 
orbital phases. In order to establish an average lightcurve, we normalized the 
data by subtracting the mean value in the case of 1999-03-10 and 14, and by 
subtracting a linear function with a slope of 0.31 mag/d for 1999-03-11 to 
account for the decline. After applying some additional minor magnitude shifts,
we then averaged the data in phase bins of 0.005 orbits. The resulting 
lightcurves are shown in Fig.\ \ref{avlc_fig}.

The shape of the lightcurve consists of two maxima and two minima of
unequal height and depth, respectively. While the double-hump structure
can be explained as being due to the ellipsoidal variation of the secondary,
the different size of the features indicates the presence of a second light
source in the system. The characteristics of the average lightcurve and their 
possible explanations are described in Table \ref{lc_tab}.

In order to test if our interpretation of the lightcurve is consistent with 
the analysis of the spectroscopic data, a double sine function was subtracted 
from the normalized combined data. The sine was shifted in phase by precisely 
0.25 orbits to correspond to the theoretical phasing of the secondary's
ellipsoidal variation, and shifted in magnitude by $+$0.04 mag from the zero 
point of the normalized lightcurve. The latter shift and the amplitude of 0.07 
mag were manually adjusted to match the shallow maximum and the deep minimum. 
Following our interpretation, these two features should be least influenced by 
the additional light source. This is supported by the fact that they 
represent the only ``orbital symmetric'' features, i.e.\ their phase difference
amounts to 1/4 of the orbital cycle. 

The residual lightcurve of this subtraction (Fig.\ \ref{lcsine_fig}) shows an 
asymmetric hump with a slow rise to a maximum at phase 0.92, and a following 
steeper decline. All this is consistent with a feature produced by a typical 
bright spot. Our rather simple approach does not really prove that this picture
is correct or represents the only possible explanation, but it indeed agrees 
well with the spectroscopically derived values (the zero phase of the orbital 
motion and the precision of the orbital period). We return to the discussion of
the lightcurve in section \ref{disc_sec}.

\section{Dynamics and Distance}

The orbital period and the spectral type of the secondary allow us to constrain
the distance and luminosity of the system. Beuermann et 
al.\ (\cite{beuermann99}) tabulate absolute magnitudes, colours, and estimated 
radii of a sample of nearby K and M dwarfs (their Table 3).  Scaling these 
data, we find that hypothetical 1 $R_{\odot}$ stars of type K4 and type M0 
should respectively have $M_V = +6$ and $+7.2$.  The orbital period and Roche
constraints (Beuermann et al.\ \cite{beuermann98}, eqn. 1) yield
\begin{displaymath}
(R_2/R_{\odot}) = f(q) 0.82 (M_2 / M_{\odot})^{(1/3)}
\end{displaymath}
at $P = 0.272$ d, where the subscript 2 refers to the secondary star, and the 
function $f(q)$ is within 3 percent of unity for $q = M_2/M_1 \le 1$.
Because the secondary is likely to be modified by mass transfer (see, e.g., 
Beuermann et al.\ \cite{beuermann98}), we do not assume a main-sequence
mass-radius relation, but rather use the evolutionary models calculated by 
Baraffe \& Kolb (\cite{baraffe00}) as an approximate guide. Interpolating
their models to 6.5 h, we find a range of masses from 0.27 to 0.81 M$_{\odot}$,
but the lowest-mass models are estimated to have spectral types near M2, which 
is later than we observe here; a more realistic range for the CW 1045+525 
secondary appears to be 0.45 -- 0.81 M$_{\odot}$.  This mass range corresponds 
to $0.76 < R_2/R_{\odot} < 0.93$, which in turn corresponds to stars 
$0.4 \pm 0.2$ mag fainter than otherwise identical stars with $R = R_{\odot}$.
Putting all this together and propogating the errors, we find the secondary 
star to have $M_V = 7.0 \pm 0.7$, where the error is dominated by uncertainty 
in the spectral type.  Taking the observed magnitude of the secondary alone as 
$V = 15.5 \pm 0.5$ (see above) then yields $m - M = 8.5 \pm 0.8$, or $d =$ 350 
-- 700 pc.  At $b = 56^{\circ}$, extinction is unlikely to be important. At its
best-estimate distance of 500 pc, CW 1045+525 would lie about 400 pc from the 
Galactic plane.

We find $K_{\rm abs} = 141 \pm 3$ km s$^{-1}$ from the MDM velocities. Without 
an eclipse to constrain the inclination, this is of somewhat limited dynamical 
usefulness, but it is of interest to check that the system parameters are 
reasonable. Assuming a white dwarf mass $M_1 = 0.7$ M$_{\odot}$ and an orbital 
inclination of $45^{\circ}$ (see below), the observed mass function implies 
$M_2 = 0.55$ M$_{\odot}$, entirely consistent with the observed spectral type.
These values are essentially arbitrary but serve to demonstrate that 
$K_{\rm abs}$ is not problematic.

\section{Discussion\label{disc_sec}}

The data presented in the previous sections, on the spectroscopic and 
photometric properties, and on the long-term behaviour, of CW 1045+525 still 
do not provide a final conclusion on the nature of this system, but they allow 
to discard or favour the following scenarios with a certain probability. 

\subsection{A nova-like?}

We regard this possibility as the least probable one. The only point in 
favour of this model is the apparent absence of outbursts. However, nova-likes
are characterized by very high mass-transfer rates, establishing a stationary,
and -- for our case important -- bright and optically thick, accretion disc
(e.g., Warner \cite{warn95}). This does neither agree with the spectroscopic
data, which show Balmer and He lines purely in emission, nor with the
photometric lightcurve, which is clearly dominated by the secondary.

There remains the possibility of a nova-like in a prolonged low state, 
classifying the system as a CV of the VY Scl type, but these stars appear to
be strongly confined to orbital periods of 3--4 h (e.g., Ritter \& Kolb 
\cite{rittkolb98}), with the sole exception of VY Scl itself ($P_{\rm orb} = 
5.6$ h; Mart\'{\i}nez-Pais et al.\ \cite{mart+00}).

\subsection{A magnetic CV?\label{polar_sec}}

In this CV-subclass, the magnetic field of the white dwarf is strong enough
to impede the formation of an accretion disc fully (in polars) or partly
(in intermediate polars). Instead, the matter stream from the secondary is
transferred along the field lines directly onto the pole(s) of the primary.
Due to the high temperatures caused by this process the higher-ionized emission
lines are usually very strong in magnetic CVs. Especially \ion{He}{II} 
$\lambda$4686 often matches H$\beta$ in strength (e.g., chapter 6 and 7 in
Warner \cite{warn95}). In the spectrum of CW 1045+525, however, this line is 
not even weakly present, and neither are any other high-ionized lines. On the 
other hand, the detected lines (Balmer and \ion{He}{I}), albeit being of only 
medium strength, are still too strong to be produced by a magnetic CV in its 
low state. 

Although the other properties would nicely fit into the scenario of a magnetic
CV in intermediate state -- no outbursts, no clear photometric or spectroscopic
evidence for an accretion disc, a dominating secondary, an additional light 
source which could be produced by an accretion column -- we consider the case 
of the missing \ion{He}{II} strong enough to assign a low probability to it.

\subsection{A dwarf nova?}

The main point which speaks against a dwarf-nova classification is the 
apparent absence of the disc: the lightcurve does not show any flickering,
there are no recorded outbursts, the emission lines do not have the
characteristic double peaks and might be produced elsewhere in the system
(e.g., in an accretion column). However, in spite of showing no flickering,
the lightcurve might actually contain evidence for the presence of a disc, in
the fact that a) the minima are of different depth, and b) that the deep
minimum occurs at the superior conjunction of the secondary. In our 
description of that phase (Table \ref{lc_tab}) we explained the deep minimum
by an at least partly obscuration of the additional light source. At superior
conjunction, the most probable component which could stand between the observer
and the combined light of the secondary and the additional source is a (faint)
accretion disc (the primary is, of course, much too small to cause the observed
broad minimum).

A faint, low-mass-transfer, accretion disc could also explain both the
apparent absence of flickering (because the variation is drowned in the 
contribution of the secondary) and of outbursts. A look into the catalogue of 
Ritter \& Kolb (\cite{rittkolb98}) shows that the outburst recurrence times for
long-period dwarf novae usually are $<$100 d, and thus rather short, but also 
that the recurrence intervals for most dwarf novae in this region are 
unknown (which might 
actually indicate long outburst recurrence times). There is also the case of 
the dwarf nova \object{CH UMa} which, at an orbital period of 8.2 h, has a 
recurrence time $\Delta t_{\rm out} \sim$300--370 d (Simon \cite{simo00}). The 
long-term monitoring of CW 1045+525 since 1999 still consists of many gaps and 
has a limiting magnitude $V \sim$14 (as shows a query in the VSNET 
database\footnote{Available on the WWW at
http://www.kusastro.kyoto-u.ac.jp/vsnet/etc/searchobs.cgi?text=CW1045\%2B525}),
so that in principle
short, small (up to $\Delta V \sim$1.5 mag) outbursts 
could have been 
occurred unnoticed.
On the other hand, a reasonable dense monitoring (50 to 125 data points
per year) in the years 1990--1999 at the automated 0.41-m RoboScope telescope
in Indiana (e.g., Ringwald et al.\ \cite{ring+96}; and references therein) 
failed to show any variations larger than 0.4 mag
(Honeycutt 2001, private communication). This lets appear the `frequent 
small-amplitude outburst' scenario rather unlikely.

If above picture applies, 
the single-peaked emission lines together with the 
photometric variation suggest a medium inclination in the range $\sim$45\degr 
-- 60\degr. The additional emission source could then be interpreted as a
bright spot and/or the mass stream. It appears probable that the additional 
continuum emission and the additional line emission do not originate from
exactly the same region, and furthermore, that (at least) two additional 
sources contribute to the emission line profile. Evidence for this comes from
the following:
\begin{itemize}
\item The phase difference of the extrema in the lightcurve, and especially of
the minima, is not exactly 0.5. This indicates that the additional light source
is located at a certain angle from the line secondary -- primary, and thus 
probably not very close to the centre of mass.
\item The diagnostic diagram (Fig.\ \ref{dd_fig}) and the radial-velocity
measurements of the line wings (Fig.\ \ref{rv_fig}) indicate additional
emission contribution also at higher velocities, as even the extreme line 
wings show a phase difference $\Delta \varphi \sim$0.1 orbits with respect to 
the absorption lines.
\item Last, not least, there is the strong emission component near the centre
of mass, which can therefore not be very much displaced from the line which
connects the primary and the secondary.
\end{itemize} 
 
While the additional continuum source and the ``high-velocity'' emission might 
well stem from the same location, which could be explained as a ``classic'' 
bright spot (a phase difference of 0.1 translates to an angle of 36\degr), the
origin of the ``stationary'' component remains unknown. Time-resolved 
high-resolution spectroscopy would thus be welcome for a detailed 
examination of the line profile and its individual components.

\section{Summary}

\begin{enumerate}
\item We have presented photometric and spectroscopic data of the cataclysmic
variable CW 1045+525.
\item The spectrum shows emission lines of the Balmer and the \ion{He}{I}
series, as well as the absorption spectrum of the late-type secondary.
\item Comparison with template spectra yielded best agreements for a
secondary spectral type of K5V -- M0V contributing $\sim$70\% of the light in
the optical range.
\item The orbital period was derived by radial-velocity measurements of the 
absorption lines of the secondary. The covered range of 495 d 
constrains the period to $P_{\rm orb}$ = 0.271278(1) d, without any
cycle-count ambiguity.
\item The emission lines have an asymmetric profile whose variations clearly
show an orbital modulation. It appears to be composed out of at least two
components: an almost stationary line centre and ``high-velocity'' wings which
move with a semi-amplitude of $\sim$130 km/s. The latter show a phase 
displacement of $\sim$0.1 orbits with respect to the absorption lines.
\item The photometric lightcurve is dominated by the ellipsoidal variation of
the secondary, but also includes light from an additional ``asymmetric'' source.
The precision of the derived period allowed an extrapolation to the photometric
data and thus an orbital phasing of the features in the lightcurve.
\item Neither the spectroscopy nor the photometry shows clear evidence for the
presence of an accretion disc, although certain lightcurve features can be
interpreted as the obscuration of the additional light source and the secondary
by a faint disc.
\item The secondary star's brightness and spectral type yield a distance
range 350 -- 700 pc.  The velocity amplitude of the absorption spectrum
does not suggest anything extraordinary about the component masses.
\item We tentatively conclude that CW 1045+525 is a long-period dwarf nova
with a faint accretion disc and 
a long outburst recurrence time.
\end{enumerate}

\begin{acknowledgements}
JRT and WHF thank the U. S. National Science Foundation for support through 
grant AST 9987334, and the MDM Observatory staff for their help. CT, NB, and LS
thank the director of the Hoher List Observatory, W.\ Seggewiss, for generous 
allocation of observing time, and the Asiago staff for their support.
We also thank R.K. Honeycutt for his helpful comments and the information
on the RoboScope monitoring.
\end{acknowledgements}

\end{document}